\documentclass[doublecol]{epl2} 
\usepackage{color}
\usepackage{graphicx}
\usepackage{amsmath, amsthm, amssymb}
\usepackage[colorlinks,citecolor=red]{hyperref}
\usepackage{braket}
\usepackage{bbold}
\definecolor{almond}{rgb}{0.94, 0.87, 0.8}
\definecolor{babyblue}{rgb}{0.54, 0.81, 0.94}
\definecolor{ashgrey}{rgb}{0.7, 0.75, 0.71}

\title{On thermalization of two-level quantum systems}
\shorttitle{On thermalization of two-level quantum systems} 

\author{Prathik Cherian J\inst{1,2} \and Sagnik Chakraborty\inst{1,2} \and Sibasish Ghosh\inst{1,2}}
\shortauthor{Prathik Cherian J \etal}

\institute{                    
  \inst{1} Optics and Quantum Information Group, The Institute of Mathematical Sciences, C. I. T. Campus, Taramani, Chennai 600113, India\\
  \inst{2} Homi Bhabha National Institute, Training School Complex, Anushakti Nagar, Mumbai 400094, India
}

\abstract{Providing the microscopic behavior of a thermalization process has always been an intriguing issue. There are several models of thermalization, which often requires interaction of the system under consideration with the microscopic constituents of the macroscopic heat bath. With an aim to simulate such a thermalization process, here we look at the thermalization of a two-level quantum system under the action of a Markovian master equation corresponding to memory-less action of a heat bath, kept at a certain temperature, using a single-qubit ancilla. A two-qubit interaction Hamiltonian ($H_{th}$, say) is then designed -- with a single-qubit thermal state as the initial state of the ancilla -- which gives rise to thermalization of the system qubit in the infinite time limit. Further, we study the general form of Hamiltonian, of which ours is a special case, and look for the conditions for thermalization to occur. We also derive a Lindblad type non-Markovian master equation for the system dynamics under the general form of system-ancilla Hamiltonian.}

\begin{document}

\maketitle

\section{Introduction}

The study of evolution of open systems towards equilibrium has always been a challenging problem in Statistical Mechanics. The difficulty lies in prescribing a form of interaction between the system and the environment at the microscopic level that will give rise to equilibration. It has been evaded by proposing the so called {\it H-theorem} which states that a system attains equilibrium when the entropy function is maximized over the accessible states of the system.

Although this has proved to be a very efficient way to calculate and work with equilibrium states, the heart of the problem remains unsolved. We look at this thermodynamic problem from a quantum mechanical perspective. Quantum Thermodynamics have received a lot attention in the recent past \cite{rt1,rt2}. The concepts and laws of thermodynamics are presumably valid only in macroscopic regime. To see how the laws and definitions of thermodynamic quantities viz heat, work, etc behave in microscopic regime is one of the main objectives of Quantum Thermodynamics.

There has been a number of works \cite{pop1,pop2,short1,short2,gogolin} where the problem of equilibration is looked at from a quantum mechanical perspective. For example, Linden et al. \cite{linden} looked into the problem of smallest possible quantum refrigerator. In the process, they considered a two-qubit system as a refrigerator in which one qubit acts as the system to be cooled while the other works as the coil of the refrigerator by extracting heat from the body (to be cooled), and releasing it to the environment. The two-qubit refrigerator is derived from the equilibrium (steady) state solution of a three-qubit master equation which the authors provided phenomenologically. This motivated us to see if, instead of following this phenomenological approach, a microscopic description for the thermalization process (equilibration to a thermal state) is possible through a thermalizing Hamiltonian. Such a simulation of the thermalization process can serve at least two purposes: (i) simulating a natural thermalization process in lab, and (ii) comparing different time scales (e.g., time scales for thermalization versus interaction time scales of different constituents of the system) without assuming a priori their ordering.

To completely characterize the joint Hamiltonian of the system and environment that results in equilibration of the system, is a formidable task. So, instead we ask the following question: whether for a given thermalization process of a system, there exists an ancilla in a specific state and a joint Hamiltonian of system-ancilla that gives rise to the exact process of equilibration on the system. In this paper, we provide an affirmative answer to this question in the case of quantum-optical master equation.

We work out a {\it thermalizing Hamiltonian} $H_{th}$ for the quantum-optical master equation \cite{text1} which gives rise to thermal equilibration of a qubit. We find that a single-qubit ancilla initialized in a thermal state is sufficient for such a dynamics to be mimicked.

Our next aim is to look for such simulations of thermalization process which evolves under the action of non-Markovian dynamics. We analyse such situations further by considering a general form of thermalizing Hamiltonian of which the quantum-optical master equation dynamics is a special case. We work out the necessary and sufficient conditions for Markovianity of the system dynamics given a form of the simulating interaction Hamiltonian. Note that not every non-Markovian dynamics gives rise to equilibration of the system, and thereby, thermalization. Our approach here provides one possible way of generating a thermalizing non-Markovian dynamics through the prescription of a simulating Hamiltonian. It is worth mentioning here that, as there are a number of definitions of Markovianity in quantum mechanical scenario \cite{markov1,markov2,markov3,markov4} we stick to the definition of completely positive (CP) divisibility \cite{markov3,markov4} and use the characterization of Wolf et. al. \cite{cirac} for finding out the aforementioned conditions.

An interesting model of thermalization was proposed by V. Scarani et. al. \cite{scarani}. Another model of thermalization (for spin-$\frac{1}{2}$ systems) has been developed by Kleinbolting and Klesse \cite{klesse}. In these works, they used the swap operation between system and bath to give rise to thermalization. But a drawback of these methods is that the system is fully thermalized after a \textit{finite} time interval, which would imply that the thermalizing map is a function of only the temperature to which the system will thermalize and the time interval taken to reach it. This proposition seems to be unrealistic as this does not take into account the intricacies of the system, environment or the correlations shared between them, that might affect the process of thermalization.

In \cite{oliveira1,oliveira2,oliveira3}, M. J. de Oliveira has shown another novel approach to thermalization for systems in contact with an environment (typically, heat reservoirs). In \cite{oliveira1}, a quantum Focker-Planck-Kramers (FPK) equation is derived via canonical quantization of the classical FPK equation to account for quantum dissipation of systems interacting with environment. The dissipation term is chosen such that the system equilibriates to the Gibbs thermal state i.e. system thermalizes. In \cite{oliveira2,oliveira3}, the quantum FPK equation is further exploited to study heat transport properties in harmonic oscillator chains and bosonic systems. Although our approach to thermalization also begins with solving a master equation, it differs from de Oliveira's in that our aim is to derive simulating Hamiltonians for thermalization and thereby study generic features of thermalization in open quantum systems.

First, we describe the thermalizing process of a qubit as a pin map. We then look at the quantum optical master equation for a qubit to find out its time dependent solutions. The affine transformation relating the initial state and the time-evolved state is then described. This affine transformation is then parametrized to find out the thermalizing Hamiltonian $H_{th}$ with a single-qubit ancilla simulating the heat bath. In the following section, we consider the thermalizing Hamiltonian in a more general form and derive the conditions on the time dependence for thermalization to occur (in the infinite time limit). We also derive the necessary and sufficient conditions for such a Hamiltonian to lead to Markovian dynamics for the system evolution. Further, we derive the Lindblad type master equation for a system dynamics arising out of our general form of Hamiltonian. And finally, we draw our conclusions.

\section{Form of thermalizing Hamiltonian}

The starting point of our work is realising that thermalization can be achieved through several ways, one of which being Markovian master equations with a thermal bath. Therefore, we take a Markovian master equation, the quantum optical master equation, where a qubit (two levels of an atom) is in contact with a bath (a system of non-interacting radiation field). Given the fact that all Markovian master equations with thermal baths give rise to equilibration to {\it thermal states} $(\rho_{th} = e^{-H/k_B T})$, with $H$ being the system Hamiltonian, we try to figure out a joint Hamiltonian between the system and an ancilla, which will do the same. This system-ancilla Hamiltonian, which will henceforth be called as thermalizing Hamiltonian $H_{th}$, will give rise to a unitary process where the system (two levels of the atom) will equilibriate to a (constant) thermal state.

To calculate the thermalizing Hamiltonian $H_{th}$, we find the affine transformation on the Bloch vector of the system qubit that will give rise to the same evolution as the quantum optical master equation. In doing so, we realize that the affine transformation is a special case of the generalized amplitude damping channel \cite{text2}. We then refer to a result by Narang and Arvind \cite{narang} where it is shown that it is enough for certain qubit channels to have a single-qubit mixed state ancilla to simulate the action of the channel as a sub-system dynamics of a system-ancilla unitary evolution. It may be noted here that Terhal et. al. \cite{smolin} have shown that certain single-qubit channels can only be simulated through qutrit mixed state environments. Incidentally, our affine transformation fits into the criterion for single-qubit ancilla as in Narang and Arvind \cite{narang}, and we find a two-qubit Hamiltonian that simulates the evolution of the system qubit via the quantum optical master equation.

Given below are the details of the aforesaid process. We will be working in the computational basis unless mentioned otherwise.

\subsection{Thermalizing maps for a qubit: Pin Map}
Before we introduce the optical master equation for a two-level quantum system, we look for the most general way a qubit can lead to thermalization -- a qubit channel -- a completely positive trace preserving map $\mathcal{N}:\mathcal{L}(\mathbb{C}^2)\rightarrow \mathcal{L}(\mathbb{C}^{2})$ such that $\mathcal{N}(\rho)=\rho_{th}=\text{diag}(p,1-p)$ for all single-qubit states $\rho$ with $0\leq p\leq 1$. Such a map is called a {\it pin} map. Here, $\mathcal{L}(\mathbb{C}^2)$ is the set of all bounded linear operators $A: \mathbb{C}^2\rightarrow\mathbb{C}^2$. Thus, we have:
\begin{equation}
\label{N}
\mathcal{N} =
\begin{bmatrix}
p & 0 & 0 & p\\
0 & 0 & 0 & 0\\
0 & 0 & 0 & 0\\
1-p & 0 & 0 & 1-p
\end{bmatrix}.
\end{equation}
The Kraus operator for $\mathcal{N}$ are:
\begin{gather*}
\label{k1}
K_{00} = \begin{bmatrix}
\sqrt{p} & 0\\
0 & 0
\end{bmatrix},
~~~
K_{01} = \begin{bmatrix}
0 & \sqrt{p} \\
0 & 0
\end{bmatrix},
\end{gather*}

\begin{gather}
\label{k2}
K_{10} = \begin{bmatrix}
0 & 0\\
\sqrt{1-p} & 0
\end{bmatrix},
~~~
K_{11} = \begin{bmatrix}
0 & 0 \\
0 & \sqrt{1-p}
\end{bmatrix}.
\end{gather}

\subsection{Optical master equation}
It would have been useful to have a dynamical version of the pin map, whose Kraus operators are given in equation (\ref{k2}). This would then give rise to a master equation corresponding to pin map, and thereby, for thermalization. In the absence of such a dynamical version in general, we now look at the optical master equation to come up with one possible dynamical version of the Kraus operators in equation (\ref{k2}).

We choose the following Markovian master equation (quantum optical master equation) which corresponds to a qubit interacting with a bosonic thermal bath under Markovian conditions.

\begin{equation}
\label{lind}
\begin{split}
\frac{d\rho (t)}{dt} &= \gamma_0 (N +1) \Big(\sigma_-\rho(t)\sigma_+ - \frac{1}{2} \{ \sigma_+\sigma_-,\rho(t) \}\Big)\\
&~~ + \gamma_0 N \Big(\sigma_+\rho(t)\sigma_- - \frac{1}{2}\{\sigma_-\sigma_+,\rho(t)\}\Big)
\end{split}
\end{equation}

Here, $N=(\exp\frac{E(\omega)}{k_B T}-1)^{-1}$ is the Planck distribution, $k_B$ is the Boltzmann constant, $T$ is temperature of the heat bath and $E(\omega)=\hbar\omega$ is the energy of the system at frequency $\omega$. $\gamma_0$ is the spontaneous emission rate of the bath, and $\gamma = \gamma_0 (2N+1)$ is the total emission rate (including thermally induced emission and absorption processes). Here we have neglected the free evolution part. For more details, refer to \cite{text1}.

If the initial system qubit state is given by $\rho(0)=\frac{1}{2}(\mathbb{1}+\bar{r}(0).\bar{\sigma})$, where $\bar{r}(0) = (r_1(0),r_2(0),r_3(0))$, the master equation can be readily solved by choosing the time-evolved state to be $\rho(t)=\frac{1}{2}(\mathbb{1}+\bar{r}(t).\bar{\sigma})$ where $\bar{r}(t) = (r_1(t),r_2(t),r_3(t))$. We find, $r_1(t) = r_1(0) e^{-\gamma t/2}, r_2(t) = r_2(0) e^{-\gamma t/2}, r_3(t) = (r_3(0)+g) e^{-\gamma t} - g.$
Here $g =\frac{\gamma_0}{\gamma}=(2N+1)^{-1}$ and so, $g\in [0,1]$. $g$ gives us a measure of the temperature $T$. It can be easily seen that, higher the value of $g$, lower the temperature and vice versa. Specifically, $g=0$ for $T=\infty$ and $g=1$  for $T=0$.
The steady state solution for the system is a thermal state as expected, and corresponds to the Bloch vector $(0,0,-g)$. Explicitly,
\begin{equation}
\label{gval}
\rho_{th} = \text{diag}(\frac{1-g}{2},\frac{1+g}{2}).
\end{equation}
\subsection{Affine Transformation}
Any single-qubit channel can be written as an affine transformation of the form $r_i(t) = \sum_{j=0}^{3}M_{ij} r_j(0) + C_i$ \cite{text2,narang}. Thus, we can express the corresponding affine transformation for our solution (given in equation (\textcolor{red}{4})) as a $3\times3$ matrix $M$ and a column matrix $C$:

\begin{equation}
\label{M1}
M =
\begin{bmatrix}
e^{-\gamma t/2} & 0 & 0\\
0 & e^{-\gamma t/2} & 0\\
0 & 0 & e^{-\gamma t}
\end{bmatrix},
C =
\begin{bmatrix}
0\\
0\\
g(e^{-\gamma t}-1)
\end{bmatrix}.
\end{equation}
Here, we notice that this affine transformation is a special kind of generalized amplitude damping channel. Amplitude damping channels describe the effect of energy dissipation to environment at finite temperature. The affine transformation for a generalized amplitude damping channel has two positive parameters $B$, $p \in [0,1]$. It is given by:
\begin{equation}
M_{GAD} =
\begin{bmatrix}
\sqrt{1-B} & 0 & 0\\
0 & \sqrt{1-B} & 0\\
0 & 0 & 1-B
\end{bmatrix},
\end{equation}
\begin{equation}
C_{GAD} =
\begin{bmatrix}
0\\
0\\
B(2p-1)
\end{bmatrix}.
\end{equation}

We can see that our thermalization process is a generalized amplitude damping channel with the parameter $p<\frac{1}{2}$.

\subsection{Parametrizing the transformation}
In \cite{narang}, Narang and Arvind used a single-qubit mixed state ancilla to parametrize the affine transformation of a single-qubit channel. We follow their technique to simulate our dynamical process for thermalization. To do so, we consider a single-qubit mixed state ancilla of the form $\rho_e=(1-\lambda)\frac{\mathbb{1}}{2}+\lambda|\phi\rangle\langle\phi|.$ where $\frac{\mathbb{1}}{2}$ is the maximally mixed state and $|\phi\rangle$ is a general pure state given by, $|\phi\rangle= \cos\Big( \frac{\xi}{2}\Big)|0\rangle+ e^{-i\eta}\sin\Big(\frac{\xi}{2}\Big)|1\rangle.$

If $\rho_e$ plays the role of a bath state of a single-qubit system then evolution through the most general two-qubit unitary $U$ (upto a freedom of local unitary actions), given in equation (\ref{ud}) below, will result in the following affine transformation  for the system qubit, as given in equations (\ref{M}) and (\ref{C}) below. Apart from $\eta,\xi,\lambda$, three more parameters $\alpha,\beta,\delta$ are needed to completely identify the channel. Thus, the class of single-qubit channels which can be simulated by a single-qubit mixed state ancilla is a six parameter family ($\alpha,\beta,\delta,\eta,\xi,\lambda$) of affine transformations:

\begin{equation}
\label{ud}
U = 
\begin{bmatrix}
\cos\frac{\alpha+\delta}{2} & 0 & 0 & i\sin\frac{\alpha+\delta}{2}\\
0 & e^{-i\beta}\cos\frac{\alpha-\delta}{2} & ie^{-i\beta}\sin\frac{\alpha-\delta}{2} & 0\\
0 & ie^{-i\beta}\sin\frac{\alpha-\delta}{2} & e^{-i\beta}\cos\frac{\alpha-\delta}{2} & 0\\
i\sin\frac{\alpha+\delta}{2} & 0 & 0 & \cos\frac{\alpha+\delta}{2}\\
\end{bmatrix}
\end{equation}

\begin{equation}
\label{C}
C =
\begin{bmatrix}
-\lambda\sin\delta\sin\beta\sin\xi\cos\eta\\
-\lambda\sin\alpha\sin\beta\sin\xi\sin\eta\\
-\lambda\sin\alpha\sin\delta\cos\xi
\end{bmatrix}.
\end{equation}
\begin{widetext}
\begin{equation}
\label{M}
M =
\begin{bmatrix}
\cos\delta\cos\beta & \lambda\cos{\delta}\sin{\beta}\cos\xi & -\lambda\sin\delta\cos\beta\sin\eta\sin\xi\\
-\lambda\cos\alpha\sin\beta\cos\xi & \cos\alpha\cos\beta & \lambda\sin\alpha\cos\beta\cos\eta\sin\xi\\
-\lambda\cos\alpha\sin\delta\sin\eta\sin\xi & -\lambda\sin\alpha\cos\delta\sin\xi\cos\eta & \cos\alpha\cos\delta
\end{bmatrix}.
\end{equation}
\end{widetext}

Refer to Appendix A for details of our calculation.

It is also important to note here that by using the ancilla qubit, we are only simulating the dynamics of the system qubit leading to the infinite time thermalization. More specifically, we do not have the ancilla state remaining static, as is the case for the bosonic bath. The ancilla state does in fact change.

Now, we compare the parametrized forms of C and M in equations (\ref{C}) and (\ref{M}) respectively with the affine transformation of quantum optical master equation in (\ref{M1}). Thus, we get a joint unitary giving rise to thermalization. One can check that the unitary does indeed lead to thermalization in the infinite time limit. Equivalently, this can also be seen by calculating the Kraus operators for the system qubit from the joint unitary operator and then applying infinite time limit, $\lim_{t\to\infty} \rho_s(t) = \rho_{th}.$

Now, the thermalizing Hamiltonian $(H_{th})$ is found and details of the derivation are given in Appendix B.
 $H_{th}$ is of the following form,
\begin{equation}
\label{hth}
H_{th}(t) = f(t) \Big(|\phi^+\rangle\langle\phi^+ | - |\phi^-\rangle\langle\phi^- |\Big),
\end{equation} 

where, 
\begin{align}
\label{ft}
&f(t)= \frac{\gamma e^{-\gamma t/2}}{2\sqrt{1- e^{-\gamma t}}},\\
&\ket{\phi^\pm} = \frac{1}{\sqrt{2}}(\ket{00} \pm \ket{11}).
\end{align}

The most general two-qubit time-dependent Hamiltonian which gives rise to the affine transformation (\ref{M1}), by acting on tensor product of arbitrary intitial state of the system qubit and the initial state of the ancilla qubit being $\rho_{th}$ (given in equation (\ref{gval})), is of the form given in equation (\ref{hth}) above.

\section{On Markovianity of Dynamics for Thermalization}

Given a 2-qubit Hamiltonian of the form,
\begin{equation}
\label{Hft}
H(t) = f(t) (\ket{\phi^+}\bra{\phi^+}-\ket{\phi^-}\bra{\phi^-})
\end{equation}
where $\ket{\phi^\pm}=(\ket{00}\pm\ket{11})/\sqrt{2}$, we can ask what are the conditions on $f(t)$ such that the system will thermalize in the asymptotic time limit. Moreover, we can ask when the evolution of the system follows Markovian dynamics. The main reason behind the search for generic properties of $f(t)$ in the above equation is to look for a generic Hamiltonian (involving ancilla) method for thermalization which does not necssarily follow from the optical master equation - in the latter case the system is known to thermalize in the infinite time limit. We can rewrite eq. (\ref{Hft}) in the Pauli basis as $f(t)(\sigma_x\otimes\sigma_x - \sigma_y\otimes\sigma_y)$. This represents a kind of spin exchange interaction similar to the double-quantum Hamiltonian used in NMR experiments \cite{dqh}.In particular, $f(t)$ can be interpreted as a time-dependent coupling strength between the spins. Such Hamiltonians can in principle be realized in lab.

\subsection{Thermalization}
Given an arbitrary initial state for the system (say, $\rho_s^i$) and an initial thermal state for the ancilla (say, $\rho_e^i=\frac{1}{2}\text{diag}(1+g,1-g)$), we can derive the condition on a generic $f(t)$ such that the system will thermalize in the infinite time limit i.e. by imposing the following constraint,

\begin{equation*}
\lim_{t\to\infty}\text{Tr}_e \Big[U(t,0) (\rho_s^i\otimes\rho_e^i)U(t,0)^{\dagger}\Big]=\text{diag}(\frac{1-g}{2},\frac{1+g}{2})
\end{equation*}
where $U(t,0)=\exp{\left(-i\int_0^t H(\tau)d\tau \right)}$ with $H(\tau)$ defined above in (\ref{Hft}) and the RHS is as we saw in (\ref{gval}).

This condition for thermalization is finally found to be,

\begin{equation}
\lim_{t\to\infty} F(t) = (2n+1)\frac{\pi}{2}
\end{equation}
where, $F(t)=\int_0^t f(\tau)d\tau$ and $n$ is any integer.
\subsection{Markovianity of System Evolution}
Another interesting question we can raise is about the nature of the system evolution under such a Hamiltonian - will it be Markovian always? To answer this we refer to \cite{cirac} in which the authors have produced necessary and sufficient conditions for a given master equation $\dot{\rho}=L_t[\rho]$ to be Markovian (CP divisible) in nature. These conditions are:

\begin{itemize}
\item $L_t$ must be hermiticity preserving.
\item $L_t^*(\mathbb{1})=0$, \text{and}
\item $\omega_c L_t^\Gamma\omega_c \geqslant 0$,
\end{itemize} 
for all times $t$, where $L_t^*$ and $L_t^\Gamma$ are the adjoint map and Choi map of $L_t$ respectively. $\omega_c=\mathbb{I}-\ket{\omega}\bra{\omega}$ is the projector onto the orthogonal complement of the maximally entangled state $\ket{\omega}=\sum_i \frac{1}{\sqrt{2}}\ket{i,i}$.

It can be seen that the hermiticity preserving condition will always be satisfied for our particular case. Imposing the other conditions, we obtain the following necessary and sufficient constraints on the time dependence of the Hamiltonian for ensuring Markovianity of the dynamical map,
\begin{align}
 0\leqslant F(t) &\leqslant \frac{\pi}{2},~\forall t\\
 \frac{d}{dt}F(t)&\geqslant 0,~\forall t 
\end{align}
Note that alternatively, we can have a monotonically decreasing $F(t)$ bounded between $[-\frac{\pi}{2},0]$ if we choose $-f(t)$ in our Hamiltonian (\ref{Hft}). 

We may now think of a functional form of $f(t)$ which satisfies the thermalization condition but violates the markovianity conditions - namely that $F(t)$ be monotonic and bounded. A simple example for such a non-Markovian thermalizing form is,

\begin{equation}
\label{eg}
F(t)=\frac{\sin(20 t)}{1+10 t}+(1-e^{-t})\frac{\pi}{2}
\end{equation}

FIG. 1 plots $F(t)=\int_0^t f(\tau)d\tau$ for $f(\tau)$ given by equation (\ref{ft}) and also $F(t)$ given by equation (\ref{eg}).

\begin{figure}
\includegraphics[scale=0.37]{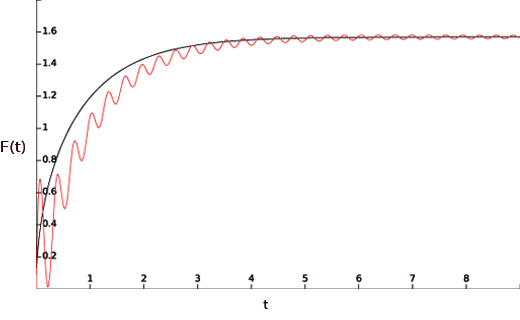}
\caption{(colour online) Red solid line is the $F(t)$ corresponding to non-Markovian thermalizing Hamiltonian while the black corresponds to that of our Markovian thermalizing form. Note that both converge to $\frac{\pi}{2}$ asymptotically and hence signify thermalization.}
\end{figure}

In the preceding sections, we have derived a specific form of thermalizing Hamiltonian from the quantum optical master equation and then we generalized it by identifying conditions for the dynamics to be Markovian. We now derive the master equation that refers to the system dynamics for thermalization under our specific form of Hamiltonian given by equation (\ref{Hft}). It is found to be of the following form,

\begin{equation}
\label{lindtype}
\begin{split}
\frac{d \rho (t)}{dt} &= \gamma_1(t) \Big(\sigma_-\rho(t)\sigma_+ - \frac{1}{2} \{ \sigma_+\sigma_-,\rho(t) \}\Big)\\
&~~ + \gamma_2(t) \Big(\sigma_+\rho(t)\sigma_-\frac{1}{2}\{\sigma_-\sigma_+,\rho(t)\}\Big)
\end{split}
\end{equation}

where,
\begin{align*} 
\gamma_1(t) &=  (1+g) f(t) \tan [F(t)] \\ 
\gamma_2(t) &=  (1-g) f(t) \tan [F(t)]
\end{align*}

Here, $F(t)=\int_0^t f(\tau)d\tau$ and $g$ is the parameter referring to the bath temperature used in defining the initial ancilla state as $\sigma_e(0)=\frac{1}{2}(\mathbb{1}+g \sigma_3)$. For more details regarding the derivation of master equation, refer to Appendix C.

The above form of master equation is immediately reminiscent of the Lindblad (Markovian) form that we have used at the beginning in equation (\ref{lind}), hence we have a master equation that is of the Lindblad type, but with time-dependent coefficients $\gamma_1(t)$ and $\gamma_2(t)$. It has been shown that the negativity of decoherence rates represent non-Markovianity \cite{hall}. Simply put, if the decoherence rates remain non-negative for all time, then the master equation represents a Markovian evolution. On the other hand, if for some time interval, it becomes negative, the dynamics is necessarily non-Markovian.

Thus, we have derived a class of master equations that can describe both Markovian as well as non-Markovian thermalization depending on the choice of $f(t)$ in the Hamiltonian. 

For example, consider the non-Markovian $F(t)$ we have defined in equation (\ref{eg}). The corresponding $f(t)$ is calculated by taking the derivative of $F(t)$ and
thus, we can derive the master equation governing such a dynamics. It can be checked that the coefficients $\gamma_1(t)$ and $\gamma_2(t)$ will not be non-negative for all time. Thus, it is seen to signify the non-Markovian nature of the dynamics.

It can also be seen that when we consider the $f(t)$ we originally derived given by equation (\ref{ft}), we recover the quantum optical master equation (\ref{lind}) with $\gamma_1(t)$ and $\gamma_2(t)$ reducing to the appropriate time-independent, positive coefficients.

\section{Conclusion}
In this paper, we look at a Markovian master equation of a qubit that leads to thermalization and simulate it through a unitary process by replacing the thermal bath with a single-qubit mixed state ancilla. Thus, we derive a thermalizing Hamiltonian for a single qubit corresponding to the quantum optical master equation. 

Although a Markovian model of thermalization has been used here, there exist non-Markovian models as well. Those models need not necessarily be simulatable through a single-qubit ancilla (mixed or pure). For example, we considered the case of post-Markovian master equation as in \cite{manis,lidar} and find that a single-qubit ancilla is not sufficient to simulate the thermalization process described therein.

We derive necessary and sufficient conditions for thermalization and Markovianity of the state evolution under a specific form (\ref{Hft}) of system-ancilla Hamiltonian. We find that it is indeed possible for us to have non-Markovian thermalization processes even for this specific kind of Hamiltonian we have described in this work.

We also derive a Lindblad type master equation for system dynamics arising out of the Hamiltonian described in our work. We see that it is possible to find signature of non-Markovian dynamics based on the negativity of decoherence rates in the master equation.

In principle, the method we have employed can be used for finding simulating Hamiltonians in higher (finite) dimensions as well. It is non-trivial because the parametrization of unitary operators for higher dimensions aren't readily available as was the case for 2-qubit unitaries. Nevertheless, in the case of infinite dimensional systems (eg: quantum harmonic oscillators), covariance matrices can be employed to proceed in this direction. As a future project, it would be interesting to study simulating thermalizing Hamiltonians for single mode harmonic oscillator.

We expect that our result will stimulate further interest in finding out the fundamental dynamics that leads to thermalization (for example, studying adiabaticity in open quantum systems). As an extension to this work, we hope to look into more general thermalization models (including non-Markovian) which will require two-qubit ancillae. Also, finding similar thermalizing Hamiltonian models for leaking cavity modes of radiation fields is an intriguing future project.

\textbf{Acknowledgements:} We would like to thank Daniel Alonso and Ramandeep Johal for insightful comments. We would also like to acknowledge productive discussions with Sandip Goyal, Manik Banik, Arindam Mallick, and George Thomas.

\section{Appendix A}
Here, we discuss the explicit calculations involved in parametrizing $C$ and $M$ matrices (appearing in equations (\ref{C}) and (\ref{M})) of the single-qubit channels simulatable through a single-qubit mixed state ancilla.

The form of $U$, given in equation (\ref{ud}), can be re-written after a simple basis change in the following way,
\begin{equation}
\begin{split}
U &= K_0(\mathbb{1}^{(s)}\otimes \mathbb{1}^{(e)}) + K_1(\sigma_1^{(s)} \otimes \sigma_1^{(e)})\\
 &+ K_2(\sigma_2^{(s)} \otimes \sigma_2^{(e)})+ K_3(\sigma_3^{(s)} \otimes \sigma_3^{(e)})
\end{split}\tag{A.1}
\end{equation}
where, 
\begin{equation}
 \begin{split}
  &K_0 = \frac{1}{2}\left(\cos\frac{\alpha+\delta}{2} + e^{-i\beta}\cos\frac{\alpha-\delta}{2}\right),\\
  &K_1 = \frac{i}{2}\left(\sin\frac{\alpha+\delta}{2} + e^{-i\beta}\sin\frac{\alpha-\delta}{2}\right),\\
  &K_2 = \frac{-i}{2}\left(\sin\frac{\alpha+\delta}{2} - e^{-i\beta}\sin\frac{\alpha-\delta}{2}\right),\\
  &K_3 = \frac{1}{2}\left(\cos\frac{\alpha+\delta}{2} - e^{-i\beta}\cos\frac{\alpha-\delta}{2}\right).
 \end{split}\tag{A.2}
\end{equation}

Now recalling the form of the mixed state ancilla $\rho_e$ and using an arbitrary initial state for the system qubit $\rho_s = \frac{1}{2}(\mathbb{1}+\bar{r}.\bar{\sigma})$, we can define the composite initial state, $\rho_{se}^{initial}=\rho_s\otimes\rho_e.$
%
%
The final time-evolved state of the system qubit can be found as, $\rho_s^{final}= \text{Tr}_e \Big[U \rho_{se}^{initial} (U)^{\dagger}\Big]$.
%

Now we can find out the components of $\rho_s^{final}$ in the basis $\{\sigma_1^{(s)},\sigma_2^{(s)},\sigma_3^{(s)}\}$ by computing Tr$[\sigma_i^{(s)}\rho_s^{final}]$. Thereby, we can read out the elements of $M$ and $C$. For example, consider $i=3$, we get:
\begin{equation}
\text{Tr}[\sigma_3^{(s)}\rho_s^{final}] = M_{31}n_1+M_{32}n_2+M_{33}n_3+C_3.\tag{A.3}
\end{equation}
Finally, 
we get the parametrized matrices $M$ and $C$ produced earlier.

\section{Appendix B}
To find the thermalizing Hamiltonian, we first need to find the values of the parameters that match with our particular case. For this, we compare the affine transformation for the quantum optical case in equations (\ref{M1}) with the parametrized matrices in equations (\ref{C}) and (\ref{M}). It can be easily seen that there exists a set of parameters as given below:
\begin{equation}
\label{set1}
\lambda=g,\cos\alpha=\cos\delta= e^{\frac{-\gamma t}{2}},\cos\beta=\pm 1 =\cos\xi,\tag{B.1}
\end{equation}
where $\eta$ can be arbitrary. So finally, we get the mixed state ancilla as the following thermal state,
\begin{equation*}
\rho_e = \frac{1+g}{2}|0\rangle\langle 0| + \frac{1-g}{2}|1\rangle\langle 1|.
\end{equation*}

Putting the values from equation (\ref{set1}) in the form of unitary given in equation (\ref{ud}), we get the unitary for the thermalization process. Note that we now have a time dependent unitary,
\begin{equation}
U(t,0) = \begin{bmatrix}
e^{\frac{-\gamma t}{2}} & 0 & 0 & i\sqrt{1- e^{-\gamma t}}\\
0 & 1 & 0 & 0\\
0 & 0 & 1 & 0\\
i\sqrt{1- e^{-\gamma t}} & 0 & 0 & e^{\frac{-\gamma t}{2}}
\end{bmatrix}.\tag{B.2}
\end{equation}
From here, we can calculate $H_{th}$ as follows. We know,
$$U(t_2,t_1) = \exp{\bigg(-i\int_{t_1}^{t_2} H(s) ds\bigg)}, \text{and}$$
\begin{equation*}
\begin{split}
U(t+\Delta t,t) &= \exp{\left(-i\int_{t}^{t+\Delta t} H(s) ds\right)}\\
&\approx \mathbb{1} -i\Delta t H(t)
\end{split}
\end{equation*}
Using the semi-group property of $U(t)$ (which holds good for small time interval $\Delta t$ even if $H$ is time-dependent) we get, 
\begin{equation*}
\begin{split}
U(t+\Delta t,0) &= U(t+\Delta t,t)U(t,0)\\		
\Rightarrow U(t+\Delta t,t) &= U(t+\Delta t,0)U^{\dagger} (t,0)\\
&= \Big(U(t,0) + \Delta t\frac{dU(t,0)}{dt}+\cdots\Big)U^{\dagger}(t,0)\\
&\approx \mathbb{1} + \Delta t \frac{dU(t,0)}{dt}U^{\dagger}(t,0)
\end{split}
\end{equation*}
Comparing with the RHS of the previous equation, we get:
\begin{equation*}
H_{th}(t) = i\left(\frac{dU(t,0)}{dt}\right)U^{\dagger}(t,0)
\end{equation*}
Thus, we get:
\begin{equation}
\begin{split}
H_{th} &= \frac{\pm \gamma e^{\frac{-\gamma t}{2}}}{2\sqrt{1-e^{-\gamma t}}} \Big(|00\rangle\langle11| + |11\rangle\langle 00|\Big)\\
&= f(t) \big(|\phi^+\rangle\langle\phi^+| - |\phi^-\rangle\langle\phi^-|\big)
\end{split} \tag{B.3}
\end{equation}
Without loss of generality, we choose the positive sign for the $f(t)$ in this paper.
%

\section{Appendix C}
We consider a Hamiltonian of the form (\ref{Hft}), with fixed initial state of ancilla qubit as $\sigma_e(0)=\frac{1}{2}(\mathbb{1}+g \sigma_3)$ (i.e. a thermal state with temperature defined through $g$ as previously explained) and an arbitrary initial state of system qubit $\rho_s(0)=\frac{1}{2}(\mathbb{1}+\bar{r}.\bar{\sigma})$ with $\bar{r}=(x,y,z)$. The time evolved state of the system under the action of such a Hamiltonian can be calculated as,
\begin{equation}
\rho_s(t)= \text{Tr}_e \Big[U(t,0) \rho_{s}(0)\otimes\sigma_e(0) (U(t,0))^{\dagger}\Big].\tag{C.1}
\end{equation}
where, $U(t,0)=\exp{\left(-i\int_0^t H(\tau)d\tau \right)}$.

Now, we use the mathematical prescription described in the Appendix of \cite{jordan} to derive the master equation for such a dynamics. First, we express $\rho_s(0)$ and $\rho_s(t)$ as vectors in the operator space of the system which has basis $\{\mathbb{1},\sigma_1,\sigma_2,\sigma_3\}$.The density matrix of the system can be represented by a $4\times1$ vector, and a superoperator on the system can be represented by a
$4\times4$ matrix. In this representation, $\boldsymbol{v}_0=\frac{1}{2}[1,x,y,z]^{T}$ is the vector form of the initial arbitrary density matrix of the system qubit and the vector form of the system qubit at time $t$ is,
\begin{equation}
\boldsymbol{v}_t=\frac{1}{2}[1,C_t x,C_t y,C_t^2 z + g S_t^2]^{T}=Q_t \boldsymbol{v}_0\tag{C.2}
\end{equation}
where, $C_t\equiv \cos(F(t)),S_t\equiv \sin(F(t))$ and $Q_t$ is the matrix representation of the system qubit evolution from the initial time to the time $t$,
\begin{equation}
Q_t=\left(\begin{array}{cccc}
1 & 0 & 0 & 0\\
0 & C_t & 0 & 0\\
0 & 0 & C_t & 0\\
g S_t^2 & 0 & 0 & C_t^2
\end{array}\right).\tag{C.3}
\end{equation}

It can be seen that $Q_t$ is invertible for finite $t$. Thus we can find that,$\partial_t\boldsymbol{v}_t=\dot{Q}_t \boldsymbol{v}_0=\dot{Q}_t Q^{-1}_t\boldsymbol{v}_t.$
Thus, $\dot{Q}_t Q^{-1}_t$ is the matrix representation of the linear transformation corresponding to the time derivative of the system density matrix, and
\begin{equation}
\dot{Q}_t Q^{-1}_t=\left(\begin{array}{cccc}
0 & 0 & 0 & 0\\
0 & \alpha_t & 0 & 0\\
0 & 0 & \alpha_t & 0\\
\beta_t & 0 & 0 & 2\alpha_t
\end{array}\right).\tag{C.4}
\end{equation}
where, $\alpha_t =-f(t)\tan(F(t)),\beta_t = 2gf(t)\tan(F(t))$.


Now we can find the superoperator corresponding to $\dot{Q}_t Q^{-1}_t$. In order to do this, we need to know the matrix representations $s_{ij}$
for the basis of the superoperator $\sigma_i[\cdot]\sigma_j$. These representations are easy to find and are given in equation (S15) in \cite{hall}. Decomposing $\dot{Q}_t Q^{-1}_t$ into the matrix representation, we get, $\dot{Q}_t Q^{-1}_t=\sum_{i,j=0}^3 a_{ij} s_{ij}.$
In our particular case, the non-zero components $a_{ij}$ turns out to be $a_{00}=4\alpha_t,a_{03}=a_{30}=\beta_t,a_{11}=a_{22}=-2\alpha_t$ and $a_{21}=-a{12}=i\beta_t$. Now by de-vectorizing, the master equation can be written as,

\begin{equation}
\begin{split}
\partial_t \rho(t) &= 4\alpha_t\rho -2\alpha_t(\sigma_1\rho\sigma_1+\sigma_2\rho\sigma_2)\\
&~~ +i\beta_t(\sigma_2\rho\sigma_1-\sigma_1\rho\sigma_2)+\beta_t\{\rho,\sigma_3\}
\end{split}\tag{C.5}
\end{equation}

Using the fact that $\sigma_\pm=\sigma_1\pm i \sigma_2$, the above equation can easily be recast into the Lindblad type master equation as given in equation (\ref{lindtype}).

\end{document}